# MeV Tau Neutrino:
## Astrophysical and Cosmological Constraints and Mischief

Geza Gyuk[a][c] and Michael S. Turner[b][c]

[a]Department of Physics,
University of Chicago, Chicago, IL  60637-1433

[b]Departments of Physics and of Astronomy & Astrophysics
Enrico Fermi Institute, The University of Chicago, Chicago, IL  60637-1433

[c]NASA/Fermilab Astrophysics Center
Fermi National Accelerator Laboratory, Batavia IL  60510-0500

Terrestrial and "Heavenly" experiments severely constrain the mass and lifetime of an MeV tau neutrino. Irrespective of decay mode, for $\tau_\nu \gtrsim 300\,\text{sec}$ the mass of the tau neutrino must be either approximately 30 MeV or less than 0.4 MeV (Majorana), 15 keV (Dirac). If the dominant decay mode includes electromagnetic daughter products, the mass must be less than 0.4 MeV (Majorana or Dirac) provided $\tau_\nu \gtrsim 2.5 \times 10^{-12}\,\text{sec}$, 15 keV (Dirac) provided $\tau_\nu \gtrsim 10^{-6}\,\text{sec}(m_\nu/\text{MeV})$. A tau neutrino of mass between 1 MeV and 30 MeV can have a host of interesting astrophysical and cosmological consequences: relaxing the big-bang nucleosynthesis bound to the baryon density and the number of neutrino species, allowing big-bang nucleosynthesis to accommodate a low (less than 22%) $^4$He mass fraction or high (greater than $10^{-4}$) deuterium abundance, improving significantly the agreement between the cold dark matter theory of structure formation and observations, and helping to explain how type II supernovae explode. Exploring the MeV mass range not only probes fundamental particle physics, but also interesting astrophysical and cosmological scenarios.

## 1. INTRODUCTION

Neutrinos are ubiquitous in the cosmos. Their relic abundance from the big bang is $113\,\text{cm}^{-3}$ (per species), and type II (core collapse) supernovae, which occur at a rate of about $10\,\text{sec}^{-1}$ in the observable Universe, produce $10^{58}$ neutrinos of energy of order 20 MeV per explosion! Even ordinary stars like our sun radiate about 3% of their power in neutrinos. Neutrinos produced in the atmosphere rain down on the earth at a rate of around one per $\text{cm}^2\,\text{sec}$.

Because of all this the heavenly lab can and has been used to obtain important constraints to the properties of neutrinos, including mass, lifetime, magnetic and transition moments, charge, velocity of propagation, secret (i.e., additional) interactions, number of neutrino types and so on [1]. These limits follow from considering (among other things): (1) the cosmic contribution of neutrinos to the mass density and through their radiative decays to the diffuse photon background (including the CBR); (2) the effects of massive neutrinos and additional neutrino species upon primordial nucleosynthesis and the formation of structure in the Universe; (3) the effects of neutrino emission upon the evolution of red giant and white dwarf stars; (4) the effects of radiative decay, mass, charge and so on on the neutrino burst from SN 1987 A detected by the Kamiokande II (K II) and Irvine-Michigan-Brookhaven (IMB) water Cherenkov detectors. A (partial) summary of the regions of the mass-lifetime plane that can be excluded is shown in Figs. 1-3.

Neutrinos can have important astrophysical and cosmological implications. For example, a neutrino species of mass 20 eV to 30 eV would account for the bulk of the mass density of the Universe, and more recently, a neutrino species of mass 5 eV to 8 eV has been suggested as an "additive" to improve the agreement between the cold dark matter picture of structure formation and observations [2]. Sciama has emphasized that a neutrino of mass 27.3 eV and radiative lifetime of around $10^{25}$ sec would explain how the bulk of the matter in the present Universe became ionized as well as accounting for the dark matter [3]. Neutrino oscillations provide a very attractive solution to the solar-neutrino problem [4] and have even been suggested as a means for explaining how supernovae explode [5].

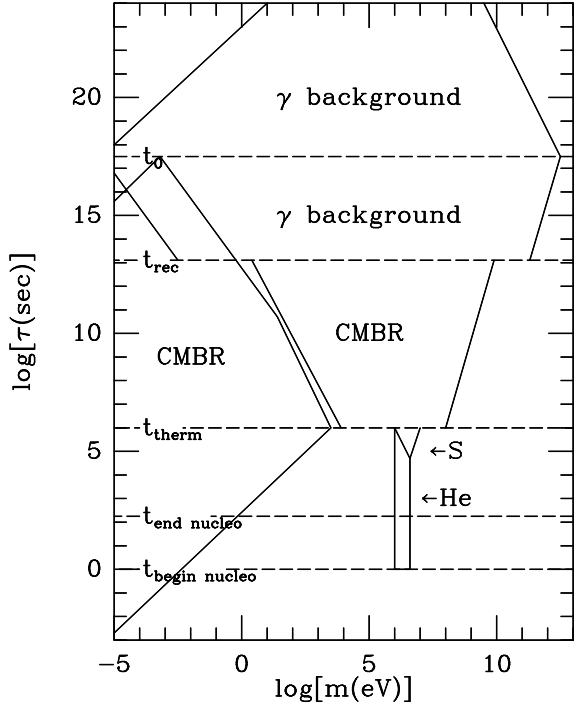

Figure 1. Excluded region of the mass-lifetime plane for a neutrino that decays radiatively. (from Ref[1])

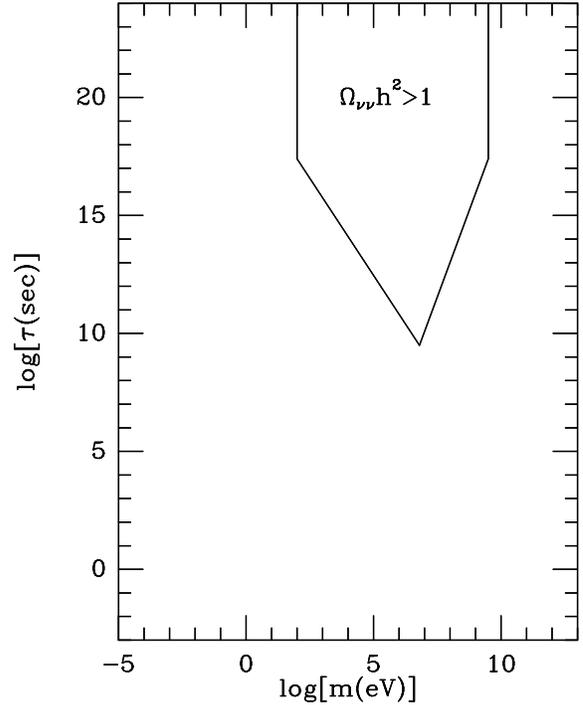

Figure 2. Excluded region of the mass-lifetime plane based upon the contribution to the cosmic mass density. (from Ref[1])

The topic of neutrinos, astrophysics, and cosmology is a very rich one indeed and it is not our intent to try to summarize it here; excellent reviews exist [1]. Rather, we will discuss recent work concerning the astrophysical and cosmological constraints to and interesting consequences of an MeV tau neutrino. This work is timely for two reasons. The current laboratory mass limit is just above 30 MeV; in the foreseeable future the tau-neutrino mass sensitivity may be improved to 10 MeV or lower. Constraints from primordial nucleosynthesis and SN 1987A allow the mass limit for a longlived ($\gtrsim$ 300 sec) tau neutrino to be lowered to around 0.4 MeV (Majorana) and to around 15 keV (Dirac). On the other hand, for masses between 1 MeV and 30 MeV there are lifetimes and decay modes that led to very interesting astrophysical and cosmological consequences: relaxing the big-bang nucleosynthesis bound to the baryon density and to the number of neutrino species, allowing big-bang nucleosynthesis to accommodate a low (less than 22%) $^4$He mass fraction or high (greater than $10^{-4}$) deuterium abundance, improving significantly the agreement between the cold dark matter theory of structure formation and observations, and help-cussing on final states containing five pions. The CLEO data set has 113 such decays and the ARGUS data set has 20 such decays. By searching for events close to the kinematic endpoint they are able to set the following 90% C.L. upper limits to the tau-neutrino mass [6]:

$$31\,\text{MeV (ARGUS)} \quad 32.6\,\text{MeV (CLEO)}. \quad (1)$$

Detector and accelerator upgrades at CLEO as well as the study of other decay modes (e.g., final states with Kaons) should lead to improved mass sensitivity, perhaps as low as 10 MeV or so. In addition, the LEP collaborations are beginning to study tau physics, including the tau-neutrino mass. Finally, upcoming experiments at B-factories (and tau/charm factories if built) may be helpful.

A beam-dump experiment at CERN using the BEBC set a very restrictive limit to the decay of tau neutrinos to channels that include electromagnetic daughter products ($e^\pm$ and photons). The absence of such electromagnetic interactions in the BEBC excludes a radiative decay rate in the interval

$$2.5 \times 10^{-12}\,\text{sec}\frac{m_\nu}{\text{MeV}} \lesssim \Gamma_{\text{rad}}^{-1} \lesssim 0.15\,\text{sec}\frac{m_\nu}{\text{MeV}}. \quad (2)$$

As we will describe, this limit together with those based upon SN 1987A and primordial nucleosynthesis all but exclude a tau neutrino that is more massive than about 0.4 MeV and that decays primarily through radiative modes.

## 2.2. SN 1987A

When the core of a massive star exhausts its nuclear fuel and collapses to form a neutron star most of its binding energy (about $3 \times 10^{53}$ erg) is released in thermal neutrinos of all three species. A neutron star is so dense that neutrinos become trapped and are emitted from a neutrinosphere whose temperature is about 7 MeV. In all, more than $10^{57}$ neutrinos (per species) of average energy around 20 MeV are emitted during the initial 5 sec to 10 sec of cooling. The detection of 19 neutrino events associated with SN 1987A by the IMB and KII detectors provided dramatic confirmation of this picture. The enormous flux of neutrinos emitted, the beautiful KII and IMB data, and our theoretical understanding of type II (core

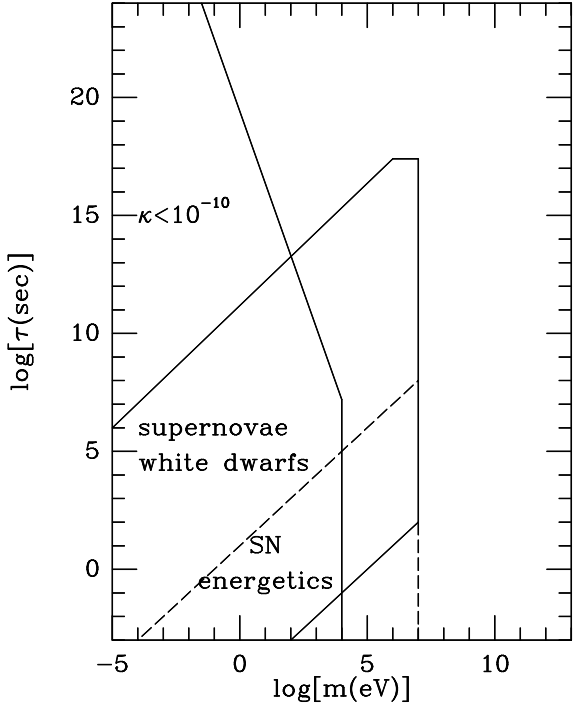

Figure 3. Excluded region of the mass-lifetime plane for a neutrino that decays radiatively based upon type II supernovae, white dwarf cooling and red giant evolution. (from Ref[1])

ing to explain how type II supernovae explode. While the theoretical motivation for an MeV-mass tau neutrino is not strong—there are some models—we wish to stress that exploring the MeV mass range allows tests of intriguing astrophysical/cosmological scenarios.

## 2. MASS/LIFETIME CONSTRAINTS

### 2.1. Laboratory

There are two very important laboratory constraints: that to the mass based upon the kinematics of tau-lepton decays and that to the radiative lifetime based on the Big European Bubble Chamber (BEBC) beam-dump experiment.

The CLEO and ARGUS collaborations have studied the decays of millions of tau leptons, fo-collapse) supernovae make SN 1987A a wonderful laboratory for probing neutrino properties, as has been summarized elsewhere [7].

So far as tau neutrinos are concerned there are three important SN 1987A constraints. The first involves Dirac neutrinos; because of the mismatch between chirality and helicity for a massive neutrino, neutrino-nucleon scattering deep inside a hot, young neutron star can transform a proper-helicity neutrino into a wrong-helicity neutrino whose interactions are weaker by a factor of $(m_\nu/2E_\nu)^2$. These wrong-helicity neutrinos are emitted copiously from the core, where temperatures reach 50 MeV or higher, and simply stream out. For a Dirac mass between about 15 keV and 1 MeV and lifetime greater than about $10^{-6}\sec(m_\nu/\text{MeV})$ they quickly rob the core of its thermal reserves leading to a burst of proper-helicity neutrinos from the neutrinosphere that is too short to be consistent with the KII and IMB data (the "timing argument") [8]. (For masses larger than around 1 MeV the wrong-helicity states become trapped and are radiated from a wrong-helicity neutrinosphere whose temperature becomes close to that of the ordinary neutrinosphere for a mass of 5 MeV; for lifetimes shorter than about $10^{-6}\sec(m_\nu/\text{MeV})$ they decay inside the neutron star.)

If Dirac $\tau$ neutrino decays produce electron or muon neutrinos or antineutrinos and their lifetime is between $10^{-6}\sec(m_\nu/\text{MeV})$ and $5 \times 10^{10}\sec(m_\nu/\text{MeV})$, then masses as low as 1 keV are excluded on the basis of the very high-energy (of order 100 MeV) events they should have produced in the KII and IMB detectors and apparently didn't [9].

The second and third constraints involve radiative decay of tau neutrinos emitted from the neutrinosphere. Decays inside the progenitor star, radius of about $3 \times 10^{12}$ cm, will be absorbed by the star and produce energy that is "visible," either thermalized and radiated from the photosphere (about $10^{49}$ erg is actually seen as the supernova fireworks) or in the bulk motion of the expanding shell (about $10^{51}$ erg is seen) [10]. Decays outside the progenitor that produce a photon lead to a flux of high-energy gamma rays that could have been seen by the SMM and PVO gamma-ray detectors which were in operation at the time. Since the neutrino fluence on earth was nearly $10^{10}$ cm$^{-2}$ and that of gamma-rays during the 10 sec interval at the time of the neutrino burst was less than about 1 cm$^{-2}$ this leads to a very stringent constraint [11]. (Additional constraints of this type have been obtained recently from GRO Comptel observations of SN 1987A at late times and of SN 1991J [12].)

Given the tau-neutrino mass, lifetime, and flux from a hot neutron star it is a simple matter to derive the constraints that follow from SN 1987A. There is a slight hitch in getting the tau-neutrino flux for masses in the MeV range: the neutrinosphere temperature is only 7 MeV (for a massless neutrino species), so that suppression of the neutrino flux should become important for masses above 10 MeV. Recently the neutrinosphere temperature and neutrino flux for a massive neutrino species has been calculated using a simple but accurate model based upon the diffusion approximation [5]. Above a mass of 10 MeV the neutrinosphere temperature slowly rises with mass, reaching about 10 MeV for a mass of 30 MeV. This means that the neutrino flux falls more slowly than a naive estimate using the Boltzmann factor for $T_\nu = 7$ MeV would suggest. The supernovae constraints based upon our fluxes for a massive tau neutrino are shown in Fig. 4.

### 2.3. Big-bang nucleosynthesis

Big-bang nucleosynthesis is one of the great successes of the standard cosmology. Provided that the baryon-to-photon ratio is between $2.5 \times 10^{-10}$ and $6 \times 10^{-10}$ the predictions for the primordial abundances of D, $^3$He, $^4$He, and $^7$Li, which span nine orders of magnitude, are consistent with their measured abundances [13]. Nonstandard assumptions about the physics of the early Universe (e.g., additional light particle species such as neutrinos, an MeV-mass tau neutrino, or a slight change in the gravitational constant) can upset this success and primordial nucleosynthesis has been used often as a heavenly laboratory to study physics beyond the standard model.

An unstable, MeV-mass tau neutrino affects nucleosynthesis in three different ways, depending upon its mass, lifetime and decay mode(s). First, the energy density of it and its daughter products contribute to the total energy density, which affects the expansion rate of the Universe. Because neutrinos cease interacting on a cosmological timescale about the time of nucleosynthesis begins ($t \sim 1\,\mathrm{sec}$), the energy density of an MeV-mass tau neutrino can exceed that of a massless neutrino species: after their annihilations freeze out the number of massive tau neutrinos remains constant so that their energy density decreases as $R^{-3}$, while that of a massless species decreases as $R^{-4}$; $R$ is the cosmic-scale factor. The main effect of the energy density is on the yield of $^4$He: higher energy density leads to more $^4$He.

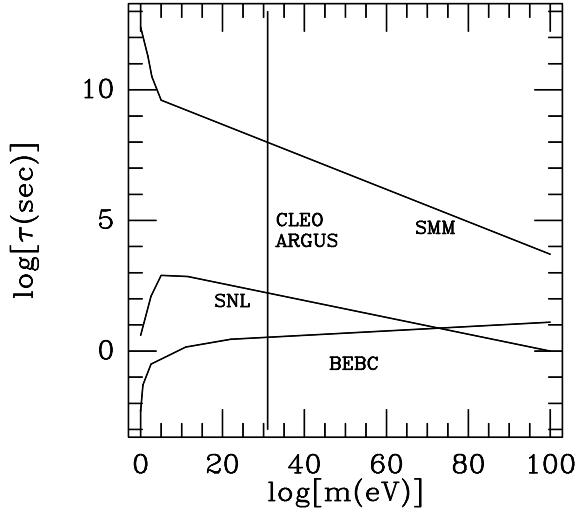

Figure 4. Excluded regions of the mass-lifetime plane for a MeV $\tau$ neutrino that decays radiatively, based upon non-detection of $\gamma$ rays from SN1987A (region labeled SMM), light and kinetic energy seen in SN1987A (below curve labeled SNL), CLEO/ARGUS mass limit, and the BEBC beam dump (below curve). These results are for a Dirac neutrino; Majorana results are similar (from Ref[5]).

Second, if the daughter products include photons or $e^{\pm}$ pairs, tau-neutrino decays produce entropy which lowers the baryon-to-photon ratio. If the decays occur around the time of nucleosynthesis, then for a fixed pre-decay baryon-to-photon the baryon-to-photon ratio at the time of nucleosynthesis is smaller, leading to decreased $^4$He production and increased D production.

The third and most interesting effect occurs if tau-neutrino decays produce electron neutrinos and antineutrinos. Through the weak interactions $n + \nu_e \leftrightarrow p + e^-$ and $n + e^+ \leftrightarrow p + \bar{\nu}_e$ these neutrinos can affect the neutron fraction, which in turn controls the amount of $^4$He synthesized (essentially all the neutrons wind up in $^4$He, so the $^4$He mass fraction produced, $Y_P \simeq 2X_n$). In the standard picture the weak interactions that regulate the neutron fraction cease occurring on a cosmological timescale when the temperature of the Universe is about 0.7 MeV; thereafter, the neutron fraction no longer tracks its equilibrium value and remains roughly constant until nucleosynthesis commences ($T \sim 0.1\,\mathrm{MeV}$), at a value $X_n \simeq 0.12$.

If decay-produced electron neutrinos and antineutrinos have "high" energies, i.e., $E_\nu \gg T, (m_n - m_p) \sim 1\,\mathrm{MeV}$, corresponding to a tau-neutrino mass greater than about 10 MeV, then the probability to convert a neutron to a proton is roughly equal to that to convert a proton to a neutron. However, there are seven times as many protons as neutrons so the net effect is to produce more neutrons than protons, increasing the neutron fraction and ultimately $^4$He production.

In the other extreme, where the decay-produced neutrinos and antineutrinos have low energies, corresponding to a tau-neutrino mass less than about 10 MeV, the conversion of protons to neutrons (but not that of neutrons to protons) is suppressed by the neutron-proton mass difference and there is a net reduction in the neutron fraction, leading to decreased $^4$He production (see Fig. 5).

We have modified the standard big-bang nucleosynthesis code to accommodate all three effects [14]. Briefly, our assumptions and changes to the code are:

1. The abundance of tau neutrinos (per comoving volume) is assumed to be constant and determined by their electroweak annihi-

lation channels, $\nu_\tau \bar{\nu}_\tau \to e^+ e^-$, $\nu_\mu \bar{\nu}_\mu$, $\nu_e \bar{\nu}_e$. The assumption that annihilations have ceased before nucleosynthesis and the neglect of inverse decays has been studied and is well justified for the masses and lifetimes of interest: $\tau_\nu/\sec \gtrsim (m_\nu/\text{MeV})^{-2}$. In Ref. [15] the regime of very short lifetimes is addressed for the decay mode $\nu_\tau \to \nu_\mu + \phi$.

2. Electromagnetic daughter products (e.g., $e^\pm$ pairs and photons) are assumed to rapidly thermalize and thereby increase the entropy density.

3. "Sterile" daughter products (i.e., those with weak interactions or weaker—muon neutrinos or Nambu-Goldstone bosons) are assumed to be relativistic and noninteracting.

4. The phase space distribution of electron and muon neutrinos and antineutrinos is followed by integrating the Boltzmann equations, including all the usual electroweak interactions ($\nu e^\pm \leftrightarrow \nu e^\pm$, $\nu\bar{\nu} \leftrightarrow e^+ e^-, \nu\bar{\nu}$, $\nu\nu \leftrightarrow \nu\nu$) as well as the decays of tau neutrinos.

5. The weak rates that control the neutron-to-proton ratio ($n + e^+ \leftrightarrow p + \bar{\nu}_e$, $n + \nu_e \leftrightarrow p + e^-$, $n \leftrightarrow p + e^- + \bar{\nu}_e$) are modified to take into account the perturbed phase-space distribution of electron neutrino and antineutrinos.

The effect of a decaying tau neutrino on primordial nucleosynthesis depends upon its decay mode. Based upon the three ways in which nucleosynthesis is affected we have identified four "generic" decay modes that bracket the larger range of possibilities:

1. Tau neutrino decays to daughter products that are "sterile," e.g., $\nu_\tau \to \nu_{\text{sterile}} + \phi$ or $\nu_\mu + \phi$ (for lifetimes greater than a few seconds it is a good approximation to treat muon neutrinos as noninteracting). For this mode the only effect on nucleosynthesis is through the energy density of the tau neutrino and its daughter products.

2. Tau neutrino decays to daughter products that include sterile particles and particles that interact electromagnetically, e.g., $\nu_\tau \to \nu_\mu + \gamma$. For this mode both the energy density of the tau neutrino and its daughter products and entropy production affect nucleosynthesis.

3. Tau neutrino decays to daughter products that include electron neutrinos, e.g., $\nu_\tau \to \nu_e + \phi$ or $\nu_e + \nu_e \bar{\nu}_e$. For this mode both the energy density of the tau neutrino and its daughter products and the change in the weak rates affect nucleosynthesis.

4. Tau neutrino decays to daughter products that include electron neutrinos and particles that interact electromagnetically, e.g., $\nu_\tau \to \nu_e + e^\pm$. For this mode all three effects come into play.

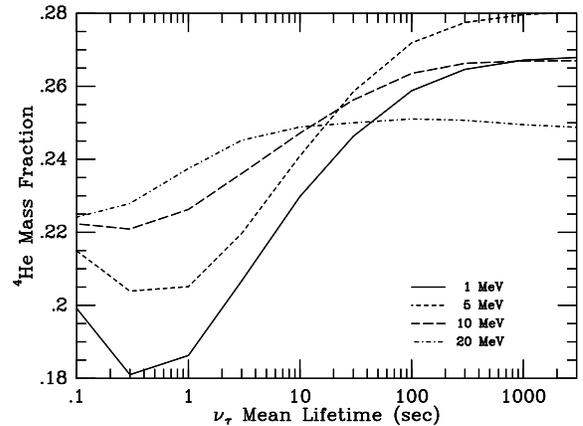

Figure 5. The effect of a $\tau$ neutrino that decays $\nu_\tau \to \nu_e + \phi$ on $^4$He production (from Ref[14]).

Figures 5 and 6 show the effect of tau neutrinos of different masses and different decay modes on the production of $^4$He as a function of lifetime. (Because the $^4$He abundance is so well known, $Y_P \simeq 0.23-0.25$ [13], it offers the most leverage.)

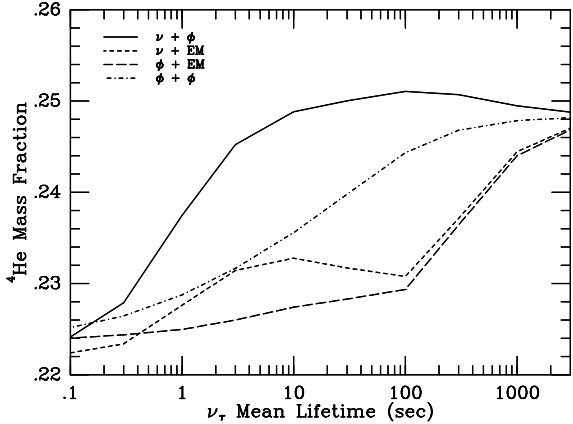
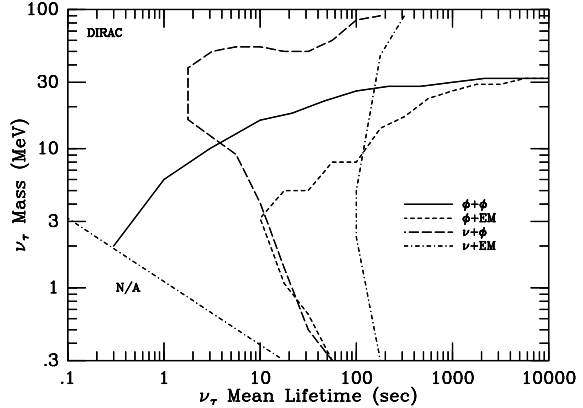

Figure 6. The effect of a $\tau$ neutrino on $^4$He production for the generic decay modes; $\nu \to \phi + \phi$ denotes the all sterile decay mode (from Ref[14]).

Figure 7. Regions of the mass-lifetime plane that are excluded on the basis of nucleosynthesis (to the right of the curves) for the four generic decay modes (Dirac). Our results are not reliable in the region denoted by N/A (from Ref.[14]).

When the effect of a decaying tau neutrino upon the yields of primordial nucleosynthesis are significant, they are almost always deleterious, and large regions of the mass-lifetime plane can be excluded. The excluded regions for the different generic decay modes are shown in Figs. 7 and 8.

### 2.4. Confluence of constraints

Bringing together all the constraints discussed above the following general statements can be made about a massive tau neutrino:

1. If the dominant decay mode is radiative and the lifetime is longer than $2.5 \times 10^{-12}\,\text{sec}(m_\nu/\text{MeV})$, the mass must be less than 0.4 MeV or less than 15 keV for a Dirac neutrino provided $\tau_\nu \gtrsim 10^{-6}\,\text{sec}(m_\nu/\text{MeV})$ [8]. In the Dirac case the lower mass limit falls to about 1 keV if the decay products include electron or muon neutrinos and $\tau_\nu \lesssim 5 \times 10^{10}\,\text{sec}(m_\nu/1\,\text{MeV})$ [9].

2. Irrespective of the decay mode, if the lifetime is longer than about 300 sec, then the mass must be either around 30 MeV or less than 0.4 MeV (Majorana), 15 keV (Dirac). As before, in the Dirac case the lower mass limit falls to about 1 keV if the decay products include electron or muon neutrinos and $\tau_\nu \lesssim 5 \times 10^{10}\,\text{sec}(m_\nu/1\,\text{MeV})$. (A mass of approximately 30 MeV is allowed as the nucleosynthesis bounds "cut out" around 30 MeV.)

3. For the specific decay mode $\nu_\tau \to \nu_\mu + \phi$, primordial nucleosynthesis has been used to exclude masses less than about 10 MeV and lifetimes less than about $10^{-2}$ sec [15].

These bounds derive in large measure from primordial nucleosynthesis, where in deriving the abundance of massive tau neutrinos at nucleosynthesis it was assumed that tau neutrinos annihilate at the rate given in the standard electroweak theory. If the tau neutrino has mass it will of course have additional interactions which could significantly enhance the annihilation cross section, reducing their abundance at nucleosynthesis and weakening the nucleosynthesis limits.

With regard to theoretical expectations for the lifetime of a massive tau neutrino; in the standard electroweak theory an MeV-mass tau neutrino can decay $\nu_\tau \to \nu_e + e^\pm$ with lifetime

$$\tau_\nu \;=\; 192\pi^3 G_F^{-2} m_\nu^{-5} \sin^{-2}\theta \cos^{-2}\theta \qquad (3)$$

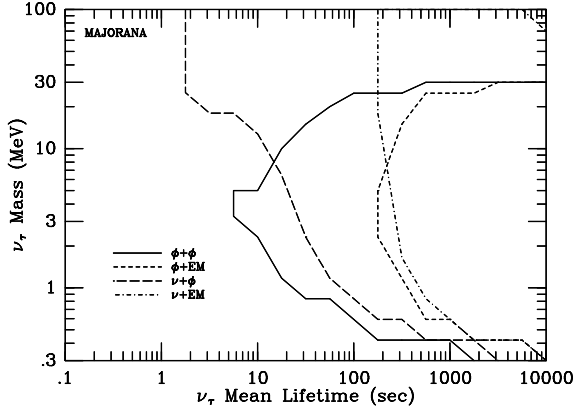

Figure 8. Same as Fig. 7 for a Majorana $\tau$ neutrino (from Ref. [14].

$$\simeq \frac{2.9 \times 10^4 \text{ sec}}{\sin^2 \theta (m_\nu / \text{MeV})^5}. \qquad (4)$$

In models with horizontal symmetries the tau neutrino can decay $\nu_\tau \to \nu' + \phi$ ($\phi$ is the Nambu-Goldstone boson of horizontal symmetry) with lifetime

$$\tau_\nu \sim 8\pi f^2 / m_\nu^3 \sim \frac{10^4 \sec (f/10^9 \text{ GeV})^2}{(m_\nu / \text{MeV})^3}, \qquad (5)$$

where $f$ is the scale of horizontal-symmetry breaking. There are other possibilities, e.g., decay mediated by righthanded gauge interactions (in this case, the lifetime in Eq. (3) scales as $M_{W_R}^4 / M_W^4$). These two examples serve to illustrate decay mediated by a massive gauge boson and scalar mediated decay.

## 3. MISCHIEF

While the astrophysical and cosmological arguments discussed above lead to very stringent and important limits, there are some very interesting islands in the mass-lifetime plane. Before describing the tantalizing astrophysical and cosmological consequences of an MeV-mass tau neutrino we wish again to disclaim any strong theoretical motivation for the masses, lifetimes, and decay modes required. Our purpose is to point out that experimentalists searching for an MeV-mass tau neutrino are also exploring interesting astrophysical and cosmological scenarios.

### 3.1. Relaxing the bound to $\Omega_B$

Big-bang nucleosynthesis constrains the contribution of baryons to the mass density of the Universe [13]:

$$0.01 h^{-2} \lesssim \Omega_B \lesssim 0.02 h^{-2}, \qquad (6)$$

where $h$ is the Hubble constant in units of $100 \text{ km s}^{-1} \text{ Mpc}^{-1}$. For $h \gtrsim 0.4$ this bounds the fraction of critical density contributed by baryons to be less than about 15%. The case for nonbaryonic dark matter hinges upon this decades-old bound, and for this reason many attempts have been made to circumvent it [17]. The most recent involved inhomogeneities in the baryon-to-photon ratio produced in a strongly first-order QCD phase transition occurring at a temperature of less than about 125 MeV. However, there is no set of parameters describing the inhomogeneity that allows the bound to be significantly loosened; moreover, current indications are that the QCD phase transition is at best weakly first-order with transition temperature 150 MeV or higher.

The upper bound to $\Omega_B$ traces to the underproduction of D and overproduction of $^4$He and $^7$Li. The overproduction of $^4$He results because for high baryon density nucleosynthesis can begin earlier, when fewer neutrons have decayed. The overproduction of $^7$Li and underproduction of D results because the neutron fraction at the time of nucleosynthesis drops precipitously for high baryon density as nuclear reactions more efficiently gobble up free neutrons [18].

Remarkably, a tau neutrino of mass 20 MeV to 30 MeV and lifetime of 300 sec to $10^4$ sec whose decays produce electron neutrinos can remedy the problems with D, $^4$He, and $^7$Li simultaneously, permitting the big-bang bound to be relaxed by a factor of ten and allowing baryons to close the Universe; see Fig. 9 [18]. It works like this. The overproduction of $^4$He is avoided because the abundance of tau neutrinos is sufficiently low that the equivalent number of massless neutrinos is about two. The D and $^7$Li problems are solved by protons capturing antielectron neutrinos which produce neutrons, preventing the neutron fraction from dropping precipitously.

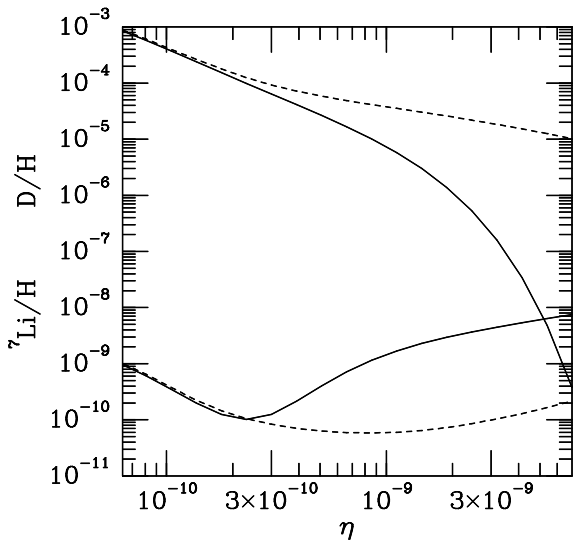

Figure 9. Deuterium and $^7$Li production as a function of baryon to photon ratio in the standard scenario (solid), and with a $\tau$ neutrino that decays $\nu_\tau \to \nu_e + \phi$ (broken) (from Ref [18]).

The loosening of the big-bang bound to $\Omega_B$ works for a wide range of tau-neutrino mass and lifetime. However, it requires that the abundance of tau neutrinos around nucleosynthesis be about a factor of ten less than the standard value, which requires that the annihilation cross section be about a factor of ten larger than that in the electroweak model. If neutrinos have mass they necessarily have additional interactions and so the annihilation cross section could well be larger.

### 3.2. Exploding supernovae

While only a small fraction (about 1%) of the energy released in a type II supernova is needed to blow up the progenitor star, creating the spectacular fireworks and preventing the formation of a black hole, numerical simulations have yet to succeed in blowing up a massive star [19]. The shock that is initiated by the core bounce stalls after traveling only a 100 km or so. Since each neutrino species carries $10^{53}$ erg "fixes" involving neutrino physics have been suggested [20]. The tricky part is getting weakly interacting neutrinos to transfer enough energy to the matter.

There is a new solution involving a 20 MeV – 30 MeV tau neutrino [5]. It works like this. Beyond the tau neutrinosphere (at a temperature of around 10 MeV) tau neutrinos continue to annihilate, producing high-energy electron, muon neutrinos, and electron-positron pairs. In total, residual tau-neutrino annihilations deposit about $10^{51}$ erg around 100 km from the core, about the right amount of energy and correct location to help power the shock; see Fig. 10. Provided the lifetime is greater than about $10^{-6}$ sec, the lifetime and decay mode are irrelevant.

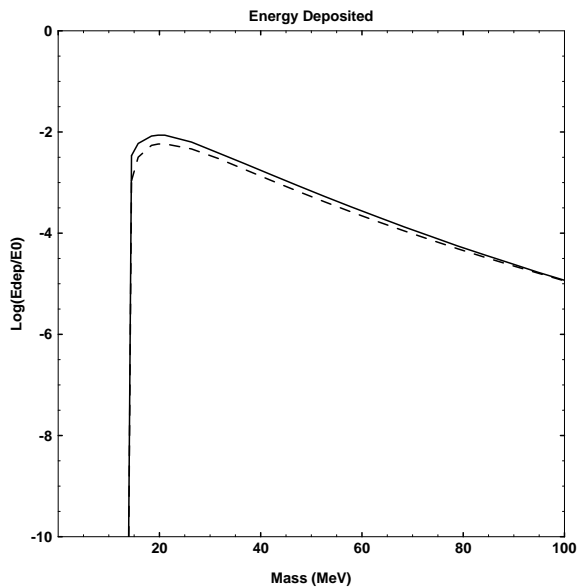

Figure 10. Energy deposited beyond the $\tau$ neutrinosphere by residual $\nu_\tau \bar\nu_\tau$ annihilations in units of $E_0 = 10^{53}$ ergs (from Ref. [5]).

### 3.3. $\tau$CDM

The cold dark matter theory of structure formation, motivated by inflation, is probably the most attractive theory of structure formation, the most studied theory of structure formation, and the most ruled out theory of structure formation. Its basic elements are: a critical Universe composed of about 5% baryons and 95% cold dark matter (slowly moving relic particles such as axions or neutralinos) with scale-invariant density perturbations. The very precise measurement of temperature fluctuations on the ten-degree scale by COBE provides the normalization for the spectrum of density perturbations.

There are now ten or so detections of CBR anisotropy spanning angular scales from about $0.5°$ to $90°$. They probe the spectrum of density perturbations on length scales from about 100 Mpc to about $10^4$ Mpc (1 Mpc corresponds to the scale of galaxies and $10^4$ Mpc corresponds to the size of the observable Universe). In addition, the distribution of matter today (more precisely, light in the form of bright galaxies) has been probed by red-shift surveys (CfA slices of the Universe, IRAS survey, APM-Stromolo survey, and others), on scales from about 1 Mpc to 300 Mpc or so. While these data confirm the general shape of the power spectrum predicted by CDM, viewed more carefully, they seem to indicate a significant problem with the simplest version of CDM: the shape of the spectrum is not quite right and the level of inhomogeneity on small scales is too high [21].

A number of variants of CDM have been proposed to remedy this problem [22]. They comprise the "CDM Family of Models:" hot (5 eV to 8 eV neutrino) + cold dark matter ($\nu$CDM) [2], tilted cold dark matter (TCDM) [23], CDM with a Hubble constant of $30\,\mathrm{km\,s^{-1}\,Mpc^{-1}}$ [24], cold dark matter + cosmological constant ($\Lambda$CDM) [25], and $\tau$CDM which involves an MeV-mass tau neutrino [26]. The last three variants rely upon the same fix: a lower ratio of matter to radiation.

While the primeval spectrum of perturbations predicted by inflation is scale invariant, (more precisely, fluctuations in the gravitational potential that are independent of scale), the spectrum of density perturbations we see today is not.

This is because the Universe underwent a transition from radiation domination at early times ($t \ll 1000$ yr) to matter domination at late times, which imposes a feature on the spectrum at the scale that crossed the horizon at matter-radiation equality (about 30 Mpc); see Fig. 11. This important scale depends upon the level of radiation in the Universe, the fraction of critical density contributed by matter (as opposed to the vacuum energy associated with a cosmological constant), and the Hubble constant (the critical density depends upon the Hubble constant). In $\tau$CDM it is the radiation level that differs from the standard scenario. The scale imposed by the transition from radiation to matter domination is roughly

$$\lambda_{\rm EQ} \sim 10 h^{-1}\,{\rm Mpc} \left(\frac{\rho_{\rm rad}}{\rho_{\rm rad-std}}\right)^{1/2} \frac{0.5}{\Omega_{\rm matter} h}. \quad (7)$$

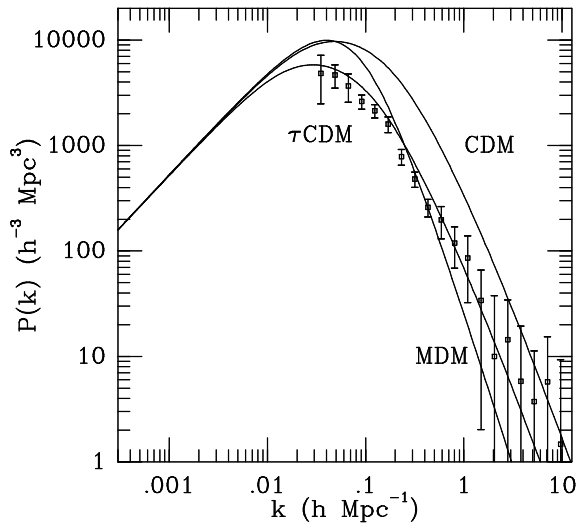

Figure 11. COBE normalized power spectra for standard CDM, MCDM and $\tau$CDM. The data points are from the IRAS 1.2 Jy red-shift survey (from Ref. [27]).

In the standard scenario the radiation content today consists of a thermal bath of photons at temperature $T_0 = 2.726 \pm 0.005\,\mathrm{K}$ and three massless (or nearly massless) neutrino species at temperature $T_\nu = (4/11)^{1/3} T_0 = 1.946\,K$. The three massless neutrino species contribute about 68% of what photons do. In order to fit the large-scale structure data better—in fact very well—the scale $\lambda_{\mathrm{EQ}}$ should be about a factor of two smaller than in the standard case, around $15 h^{-1}\,\mathrm{Mpc}$ to $20 h^{-1}\,\mathrm{Mpc}$. This can be accomplished by decreasing $\Omega_{\mathrm{matter}} h$ by a factor of $0.5 - 0.66$ (which can be done with a lower Hubble constant or smaller matter density) or by increasing $\rho_{\mathrm{rad}}$ by a factor of $2.25 - 4$. In terms of additional light neutrino species the latter corresponds to $\Delta N_\nu = 6-20$, which is clearly ruled out by big-bang nucleosynthesis (see below) and measurements of the $Z$ resonance which imply that $N_\nu = 2.99 \pm 0.02$.

An MeV-mass tau neutrino can lead to increased radiation without violating either bound! Suppose the tau neutrino has a mass of between $1\,\mathrm{MeV}$ and $10\,\mathrm{MeV}$, decays with electron neutrinos as daughter products, and has a lifetime of around $10\,\mathrm{sec}$ to $60\,\mathrm{sec}$. Tau neutrino decays do two things: First, they produce additional electron neutrinos (and possibly other relativistic particles); by virtue of the fact that the decays occur when the tau neutrino is very nonrelativistic the energy density produced is equivalent to many neutrino species, thereby raising the energy density in radiation by the required amount. Second, the electron neutrinos produced depress the neutron fraction and ultimately the $^4$He abundance, thus preventing overproduction of $^4$He that would results from the higher energy density.

### 3.4. Relaxing the bound to $N_\nu$

The constraint to the number of light (mass less than about $1\,\mathrm{MeV}$) particle species based upon big-bang nucleosynthesis is probably the best known of all the important astrophysical and cosmological limits. Expressed in equivalent number of neutrino species the limit is: $N_\nu \leq 3.4$ [27]. The limit is based upon the overproduction of $^4$He; additional light particle species lead to an increase in the energy density (at fixed temperature), in turn leading to more expansion. This leads to an earlier freeze of the neutron fraction, at a higher value, and thereby to more $^4$He production.

As just mentioned, a tau neutrino that decays and produces electron neutrinos around or shortly after the neutron fraction freezes out depresses the neutron fraction and $^4$He production, thereby making room for additional light particle species. The effect can be enormous: the equivalent of 16 additional neutrino species can be tolerated without overproducing $^4$He (see Fig. 12). Relaxing this big-bang limit could resurrect interesting particle-physics theories that were discarded because they predict too many additional light degrees of freedom.

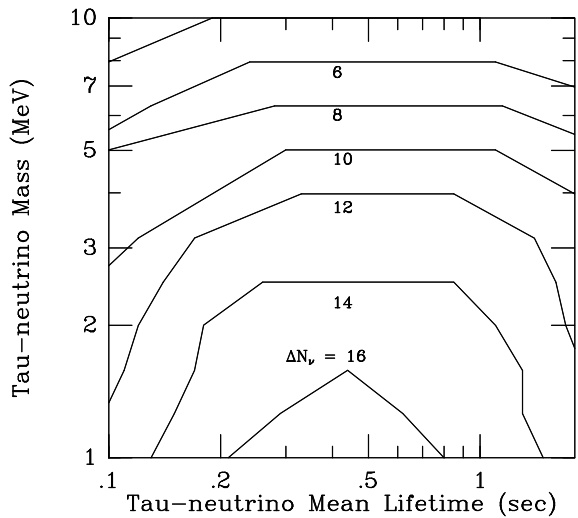

Figure 12. Additional massless species (expressed in equivalent number of massless neutrino species) permitted for the $\nu_\tau \rightarrow 3\nu_e$ decay mode (from Ref. [27]).

### 3.5. Saving big-bang nucleosynthesis itself

The agreement between the predicted light-element abundances and their measured abundances is perhaps the most stringent test of the standard cosmology. The agreement has become more impressive with time: Shortly after the dis-covery of the CBR the main success was the explanation of the large primeval $^4$He abundance; by the mid 1970's it was realized that the big-bang was the only plausible source for D; and in the 1980's both $^3$He and $^7$Li were added to the list of successes.

At present big-bang nucleosynthesis can account for the measured primordial abundances of all four light elements provided that the baryon-to-photon ratio lies in a very narrow interval: $2.5 \times 10^{-10} \leq \eta \leq 6 \times 10^{-10}$ [13]. With time the concordance interval has shrunk—and could even disappear(!). For example, should the primeval $^4$He abundance be shown to be 22% or less, $^4$He would push $\eta$ out of the interval required for the other three elements; some have argued that the primeval $^4$He abundance is this small [28]. Likewise, the recent *tentative* detection of deuterium in a hydrogen cloud at red shift $z = 3.32$, seen in absorption in the spectrum of a QSO at red shift $z = 3.42$, forces $\eta$ outside the aforementioned concordance interval [29]. At the moment, neither poses a serious threat to the standard picture. However, should that change, either could be explained by an MeV-mass tau neutrino! (Either could also be explained by a change in our understanding of the chemical evolution of $^3$He. The lower limit to $\eta$ is based upon the overproduction of D + $^3$He, and hinges upon the fact that *known* stars cannot efficiently destroy $^3$He [30]. If this argument is wrong, then the lower bound to $\eta$ becomes less stringent and low $^4$He or high deuterium could be accommodated.)

As previously mentioned, a tau neutrino of mass 1 MeV to 10 MeV and lifetime 0.1 sec to 100 sec whose decays produce electron neutrinos can depress $^4$He production. Likewise, a tau neutrino of mass 20 MeV to 30 MeV and lifetime 300 sec to $10^4$ sec whose decays produce electron neutrinos can enhance D production, even for large values of $\eta$. While it is unlikely that big-bang nucleosynthesis will need such assistance, an MeV tau neutrino could provide it.

## Acknowledgments

This work was supported in part by the Department of Energy (at Chicago and Fermilab) and by the NASA through grant NAGW-2381 (at Fermilab). GG was supported by an NSF predoctoral fellowship.